\documentclass[11pt]{article}
\usepackage{verbatim}
\usepackage[affil-it]{authblk}
\usepackage[utf8]{inputenc}
\usepackage{graphicx}
\usepackage[margin=1.25in]{geometry}
\usepackage[usenames,dvipsnames]{color}
\usepackage{url}
\usepackage[colorlinks = true,
            linkcolor = blue,
            urlcolor  = blue,
            citecolor = blue,
            anchorcolor = blue]{hyperref}

\newcommand\pubnumber{IF5: Micro Pattern Gas Detectors (MPGDs)}
\newcommand\pubdate{\today}

\title{\bf MPGDs for TPCs at future lepton colliders}

\author[1]{Alain Bellerive \thanks{Coordinator email: AlainBellerive@carleton.ca}}
\author[2]{on behalf of the LCTPC Collaboration}
\author[3]{Jochen Kaminski}
\author[3]{Peter M. Lewis}
\author[4]{Paul Colas}
\author[5]{Ralf Diener}
\author[6]{Peter Kluit}
\author[7]{Ronald Dean Settles} 
\author[6]{Jan Timmermans}
\author[4]{Maxim Titov}
\author[3]{Andreas L\"{o}schcke Centeno}
\author[3]{Christian Wessel}
\author[8]{Oskar Hartbrich}
\author[8]{Sven Vahsen}
\author[9]{Carlos Mari\~{n}as}
\author[10]{Huiron Qi}
\author[11]{Zhiyong Zhang}

\affil[1]{\normalsize Department of Physics, Carleton University, Ottawa, ON, K1S 5B6, Canada}
\affil[2]{\normalsize https://www.lctpc.org/e9/e57037/}
\affil[3]{\normalsize University of Bonn, Institute of Physics, Nu\ss{}allee 12, 53115 Bonn, Germany}
\affil[4]{\normalsize CEA/Irfu, Universit\'{e} Paris Saclay, F91191 Gif sur Yvette cedex, France}
\affil[5]{\normalsize Deutsches Elektronen-Synchrotron DESY, A Research Centre of the Helmholtz Association, 22603 Hamburg, Germany}
\affil[6]{\normalsize Nikhef, National Institute for Subatomic Physics, 1009 DB Amsterdam, Netherlands}
\affil[7]{\normalsize Max-Planck-Institut fur Physik (Werner-Heisenberg-Institut), 80805 M\"{u}nchen, Germany}
\affil[8]{\normalsize University of Hawaii, Department of Physics and Astronomy, Honolulu, HI 96822, USA}
\affil[9]{\normalsize University of Valencia - CSIC, Instituto de Fisica Corpuscular (IFIC), Spain}
\affil[10]{\normalsize Institute of High Energy Physics, Chinese Academy of Sciences, Beijing 100049, China}
\affil[11]{\normalsize University of Science and Technology of China, Hefei 230026, China}

\newcommand\pubblock{\rightline{\begin{tabular}{l} \pubnumber\\
         \pubdate \end{tabular}}}
\newenvironment{Abstract}{\begin{quotation} \begin{center}
                       ABSTRACT
     \end{center}\bigskip  }{\end{quotation}}

\def\beq{\begin{equation}}
\def\eeq#1{\label{#1}\end{equation}}
\def\eeqn{\end{equation}}

\newenvironment{Eqnarray}%
   {\arraycolsep 0.14em\begin{eqnarray}}{\end{eqnarray}}
\def\beqa{\begin{Eqnarray}}
\def\eeqa#1{\label{#1}\end{Eqnarray}}
\def\eeqan{\end{Eqnarray}}

\let\bar=\overbar

\def\lsim{\mathrel{\raise.3ex\hbox{$<$\kern-.75em\lower1ex\hbox{$\sim$}}}}
\def\gsim{\mathrel{\raise.3ex\hbox{$>$\kern-.75em\lower1ex\hbox{$\sim$}}}}

\def\del{\partial}
\def\Dslash{\not{\hbox{\kern-4pt $D$}}}
\def\dslash{\not{\hbox{\kern-2pt $\del$}}}
\def\pslash{\not{\hbox{\kern-2pt $p$}}}
\def\ETmiss{\not{\hbox{\kern-4pt $E$}}_T}

\def\Dlr{\mathrel{\raise1.5ex\hbox{$\leftrightarrow$\kern-1em\lower1.5ex\hbox{$D$}}}}

\def\MSB{{\bar{M \kern -2pt S}}}
\def\msb{{\bar{\scriptsize M \kern -1pt S}}}

\def\drb{{\bar{\scriptsize D \kern -1pt R}}}

\newcommand\snowmass{\begin{center}\rule[-0.2in]{\hsize}{0.01in}\\\rule{\hsize}{0.01in}\\
\vskip 0.1in Submitted to the  Proceedings of the US Community Study\\ 
on the Future of Particle Physics (Snowmass 2021)\\ 
\rule{\hsize}{0.01in}\\\rule[+0.2in]{\hsize}{0.01in} \end{center}}

\begin{document}

\maketitle

 \begin{Abstract}
\noindent 
This submission will focus on advancements and advantages of Micro Pattern Gas Detector (MPGD) technologies and their applications to the construction 
of a dedicated Time Projection Chamber (TPC) that can serve as an excellent main tracker for any multipurpose detector that can be foreseen to operate at a future lepton collider. 
The first portion of the report will be the executive  summary. It will be followed by sections detailing the applications of MPGDs specifically to the construction of the LCTPC for the ILD at ILC,
for a possible upgrade of the Belle II detector at SuperKEKB and for the design of a TPC for a detector at CEPC. MPGD technologies offer synergies with other detector R\&D and several application domains; 
a few examples will be provided in the context of the ongoing Snowmass long range planning exercise in the USA. Links to industrial partnership and work with institutions in the USA will be highlighted when appropriate.
\end{Abstract}

\def\thefootnote{\fnsymbol{footnote}}
\setcounter{footnote}{0}

\newpage

\pubblock

\snowmass

\section{Executive Summary}

Advances in our knowledge of the structure of matter during the past century have been made possible largely 
through the development of successive generations of high energy particle accelerators, as well as a continued improvement in detector technologies.
The physics goals of future high-luminosity lepton colliders, at the energy-frontier for the deployment of a Higgs factory and also
at the flavour-precision frontier, have put stringent constraints on the need to develop novel instrumentation. Time Projection Chambers (TPCs) operating at $e^+e^-$ machines 
in the 1990's reached their sensitivity limit and new
approaches needed to be developed to overcome the need for improved resolution. The spatial and timing resolution goals nowadays represent an order of magnitude improvement 
over the conventional proportional wire/cathode pad TPC performance, which is limited by the intrinsic $\bf{E} \times \bf{B}$ effect near the wires, and approaches the fundamental limit imposed by diffusion. 
Other detrimental effects such as material budget, cost per readout channel and power consumption also represent serious challenges for future high-precision tracking detectors. 
One of the most promising areas of R\&D in subatomic physics is the novel development of gaseous detectors. Micro Pattern Gas Detector (MPGD) technologies have become a well-established 
advancement in the deployment of gaseous detectors because those will always remain the primary choice whenever large-area coverage with low material budget is required for particle detection.  
MPGDs have indeed a small material budget,  which is important in a high background or a high-multiplicity environment, and naturally reduce space charge build up in the drift volume by 
suppressing positive ion feedback from the amplification region. Of greatest importance however, is that the $\bf{E} \times \bf{B}$ effect is negligible for an MPGD because the micro holes 
have $\sim$100 $\mu$m spacing, which offers a rotationally symmetric distribution and thus no preferred track angle. 

MPGDs, in particular the Gas Electron Multiplier (GEM),  the Micro-Mesh Gaseous Structure (Micromegas, or MM), GridPix, and other micro pattern detector schemes, offers the potential to deploy new gaseous 
detectors with unprecedented spatial resolution, high rate capability, large sensitive area, operational stability and radiation hardness.
Many detector designs aimed at future lepton colliders utilize MPGD devices. This report mainly focuses on future proposed MPGD-based TPCs at lepton colliders. 
Namely: (i) the International Large Detector (ILD) at the International Linear Collider (ILC), (ii) the Belle II detector upgrade at the SuperKEK B-Factory, and (iii) the TPC for a detector at 
the Circular Electron Positron Collider (CEPC). 

Overall, an MPGD-based TPC offers excellent tracking ability, while enabling continuous or power-cycled readouts. Historically, TPCs were the main central tracking 
chambers of ALEPH and DELPHI at the electron-positron collider LEP, where Americans were collaborators. The T2K  Near Detector with Micromegas represents another 
area where TPC technology was deployed with engagement of participants from North America. The upgrade of the ALICE TPC is a more recent example of the usage of MPGDs 
with participation from institutions from the United States. The ALICE main central-barrel tracking used to rely on multi-wire proportional chambers, 
which have since been replaced by a TPC with GEM readouts designed in an optimized multilayer configuration, which stand  up to the technological challenges imposed by continuous TPC operation at high rates. The requirement to keep the ion-induced space-charge distortions at a tolerable level, which leads to an upper limit of 2\% for the fractional ion backflow, has been achieved. The
upgraded TPC readout will allow ALICE to record the information of all tracks produced in 
lead-lead collisions at rates of 50 kHz, while producing data at a staggering rate of 3.5 TB/s. For both the T2K and ALICE TPCs, partnership with CERN allows the fabrication of 
anode boards of size of order of 50 cm x 50 cm. 

The TPC concept is viewed in particle physics as the ultimate drift chamber since it provides 3D precision tracking with low material budget and enables particle identification through dE/dx 
measurements with cluster counting techniques. At ILC and CEPC, as well as for Belle II upgrades, MPGD TPC technologies are the preferred main tracking system for some conceptual
detectors. There are synergies with other MPGD detector activities (as summarized here) that offer
clear motivation for gaseous tracking at lepton colliders. Gaseous tracking devices have been extremely successful in providing precision 
pattern recognition. They provide hundreds of measurements on a single track, with an extremely low material budget in the central region of the detector. This results in accurate 
track reconstruction and hence high tracking efficiency. The continuous measurements of charged particle tracks allow for precise particle identification capabilities, which have the possibility not
only to achieve excellent continuous tracking, but also to improve jet energy resolution and flavour-tagging capability for an experiment at a future lepton collider. These are two essential 
advantages for experiments at a lepton collider. The main challenges for the design of a 
large TPC are related to the relative high magnetic field , in which some foreseen detectors are planned to operate. For accurate measurements of the 
momenta of charged particles, the electromagnetic field has to be known with high precision. Final and sufficient calibration of the field map can be achieved using corrections derived 
from the events themselves, or from dedicated point-like and line sources of photoelectrons produced by targets
located on the end-plates when illuminated by laser systems. While the event rate at lepton collider detectors can easily be accommodated by current TPC readout technology, R\&D to mitigate the effects of secondary processes from bunch-bunch interactions is ongoing. MPGD technologies offer a wide-range of applications and call for synergy in detector R\&D at future 
lepton colliders. The availability of a highly integrated 
amplification system with readout electronics allows for the design of gas-detector systems with channel densities comparable to that of modern silicon detectors. 
This synergy with silicon detector ASIC development is very appealing for MPGD TPCs since recent wafer post-processing enables the integration of gas-amplification 
structures directly on top of a pixelized readout chip. 

The ILD TPC is in fact based on mature hardware and software contributions from multiple partners
and in particular from the United States ($e.g.$ Cornell University and Wilson Laboratory - now the Cornell Laboratory for  Accelerator-Based Sciences and Education).
The LCTPC is conceptually ready as it meets design specifications and is engineeringly possible. It is the outcome of decades of research and innovation in MPGDs.
Single-hit transverse resolution results from testbeam at 1 T magnetic field extrapolated
to the 3.5 T field of ILD clearly demonstrate that single point resolution of 100 $\mu$m after $\sim$2 m of drift over about 200 measurement points 
is achievable with several MPGD technologies (GEM, MM or GridPix). LCTPC achieved unprecedented spatial resolution of 35 $\mu$m at zero drift distance for 2 mm wide readout pads, a world record, and 55 $\mu$m with 3 mm wider pads in a high field magnet~\cite{dixit}. This translates from both simulations and measurements to two-hit separation of $\sim$2mm and a momentum resolution of $\delta(1/p_T) \simeq 10^{-4} /$ GeV/c (at 3.5 T), which are 
the required performance of the TPC as a standalone tracker at ILD for ILC. Other areas of MPGD developments are ongoing on ion gating, dE/dx, power-pulsed electronics and cooling. Similar simulations were performed by members of the Belle II Collaboration showing 
that a GridPix-based TPC could be the ultimate central tracker for an upgrade detector at a future ultra-high luminosity B-Factory. The readout choice will need to be adapted to the
beam structure of an ultra-high luminosity SuperKEKB upgrade, and probably a buffer that can handle discrete readout of multiple concurrent events will be required.
The baseline design of a CEPC detector is an ILD-like concept, with a 
superconducting solenoid of 3.0 Tesla (Higgs run) and 2.0 Tesla (Z pole run) surrounding the inner silicon detector, the TPC tracking
detector and the calorimetry system.
The CEPC TPC detector will operate in continuous mode on
the circular machine. As for the ILD TPC, MPGD technologies are applicable and desirable for a detector at CEPC.


\section{LCTPC for ILD at ILC}

The International Large Detector (ILD) is one of two proposed all-purpose detectors for the
future International Linear Collider (ILC). To meet the stringent resolution requirements for a
detailed exploration of the physics at the TeV scale, the ILD proposes a gaseous detector as the central tracking. 
Considerable R\&D on novel gaseous detectors for the ILC has been carried out during the last two decades. The ILC is the 
most advanced concept-ready accelerator to be deployed as a Higgs Factory. The detector
technologies associated with ILC are quite mature~\cite{ILC:TDR,ILC:Global}. The ILD is one detector concept at the ILC where calorimetry and tracking systems are combined~\cite{ILD:paper}. The tracking system consists of a silicon inner vertex detector, forward tracking silicon-based disks and a large volume Time Projection Chamber (TPC). A TPC using gaseous MPGD technology is being 
planned for ILD. The ILD TPC will fill a large volume about 4.7 m in length, spanning radii from 
33 cm to 180 cm (at 3.5 T) or 143 cm (at 4 T). In this volume the TPC provides up to 
220 three-dimensional points (even more for GridPix) for 
continuous track reconstruction. This high number of points allows for a reconstruction of the charged particle components of the event with high accuracy, including the reconstruction of
secondaries, long lived particles, and potentially kinks. 
The ILD TPC requires transverse ($r-\phi$) and longitudinal ($z$) single-hit space-point resolutions of less than 100~$\mu$m and 1400~$\mu$m, respectively,
for all tracks over a 2.1 m drift region. The readout electronics for the TPC has to be adapted to the design of the tracking chamber and the beam structure of the collider. The physics 
goals drive track reconstruction resolution and MPGD pad sizes as the ILD TPC requires momentum resolution $\delta(1/p_T) \simeq 10^{-4} /$ GeV/c (at 3.5 T), dE/dx resolution 
of about 5\% or better, and two-track separation of $\sim$2 mm and $\sim$6 mm in the $r-\phi$ 
and $z$ planes, respectively. A tracking efficiency of greater than 99\%, for track momenta above 100 MeV/c within the angular acceptance 
was proven to be achievable with events simulated realistically with full backgrounds. At the
same time the complete TPC system will introduce only about 5\% of a radiation length into the ILD barrel allowing particle flow algorithm technique for global event reconstruction.

Within the framework of the LCTPC Collaboration~\cite{LCTPC:refs}, a large prototype TPC has been built as a demonstrator. Its endplate can accommodate up to seven modules of MPGD, 
representative of the near-final proposed design of the TPC endplate for ILD and reconstruction of hits over a track-length of about 70 cm. 
LCTPC is a collaboration of physicists, engineers, technicians, students and support staff from 25 institutes from 12 countries with 23 other institutes as observers.
The LCTPC observer institutes for the USA are: Iowa State University, MIT, Purdue University, Yale University, 
Cornell University, Indiana University, Stony Brook, Louisiana Tech, LBNL and BNL. The MPGD technologies being developed for the LCTPC are Gas Electron Multiplier (GEM), Micromegas (MM) 
and GridPix. All technologies have been studied with an electron beam in a 1 Tesla magnet at DESY. Successful test beam campaigns with multiple modules of MPGD readouts have been carried 
out in the last few years. Major advancements have indeed been accomplished by the LCTPC Collaboration to establish MPGDs as a solid baseline for a TPC at ILC. Results demonstrate that 
the required hit reconstruction efficiency, electric field inhomogeneity, spatial resolution and stand alone momentum resolution are achievable. 

TPCs have been successfully deployed at LEP in the 1990's. The advantages of a TPC at lepton colliders are its ability to reconstruct track from charged particle in 3 dimensions, while introducing very 
small amounts of dead material. It allowed for powerful way to perform continuous pattern recognition with precise energy lost measurement for particle reconstruction and identification, 
respectively. A limitation of conventional wire-based TPCs that operated at LEP was the appearance of some field distortions due to $\bf{E} \times \bf{B}$ near the wires and field inhomogeneity 
created by ion backflow.  It is clear that the advantages of the MPGDs were promptly acknowledged with the ion backflow being very limited by a suitable choice of the field configuration, 
as well as the $\bf{E} \times \bf{B}$ effects being nearly eliminated with the microscopic structure of a MPGD. However, it was also recognized that, to profit from the
excellent resolution allowed by a limited diffusion and a very localized avalanche, either sufficiently small pads would be needed, to share the charge among several pads, or a mechanism for 
spreading the avalanche was needed. Without such a sharing, the only information obtained would have been which pad received the charge, and the hit position would have a flat probability 
over the pad width, limiting the resolution along a pad row to be $w/\sqrt{12}$, $w$ being the pitch-width over a pad row. Thus, multi-stage GEM were developed to allow for natural spreading by 
diffusion in the multilayer gas 
amplification itself, where about $\sim$300 $\mu$m r.m.s., was sufficient to obtain enough charge spreading with $\sim$1 mm wide pads. For Micromegas, where the avalanche has 
typically a $\sim$15 $\mu$m r.m.s., an additional charge-spreading mechanism was necessary. A resistive layer using a superposition of an insulator and a resistive cover provides a continuous 
Resistor-Capacitance (RC) network over the surface which spreads the charge around the avalanche. Such construction technique is applicable to MM (and GEM) to allow pad widths of 2-3 mm.
The pixel-based TPC, where single primary ionization can be detected, is now a realistic option for ILD. To make the most of the fine pitches of an MPGD, the GridPix readout structure is 
adapted to the same feature size. Readout ASICs of silicon pixel detectors, such as the Timepix3 chip, are placed directly below the gas amplification stage. In this setup, the bump bond pads 
normally used to connect the readout chip to the silicon-sensor are used as the charge collection anodes. Such principal can be applied to triple-GEM or MM. The latter MM 
incarnation is referred to as GridPix and is 
produced by using a post-processing technique, which guarantees a high quality grid perfectly aligned with the readout pixels. This alignment ensures that the complete charge avalanche initiated by a primary electron within the MM gap is collected on one pixel. Because of the high signal-to-noise ratio tracking and dE/dx measurement with GridPix both benefit from distinguishing and
 detecting single primary electrons with a high efficiency.

\begin{figure}
\begin{center}
\includegraphics[angle=0,width=0.9\textwidth]{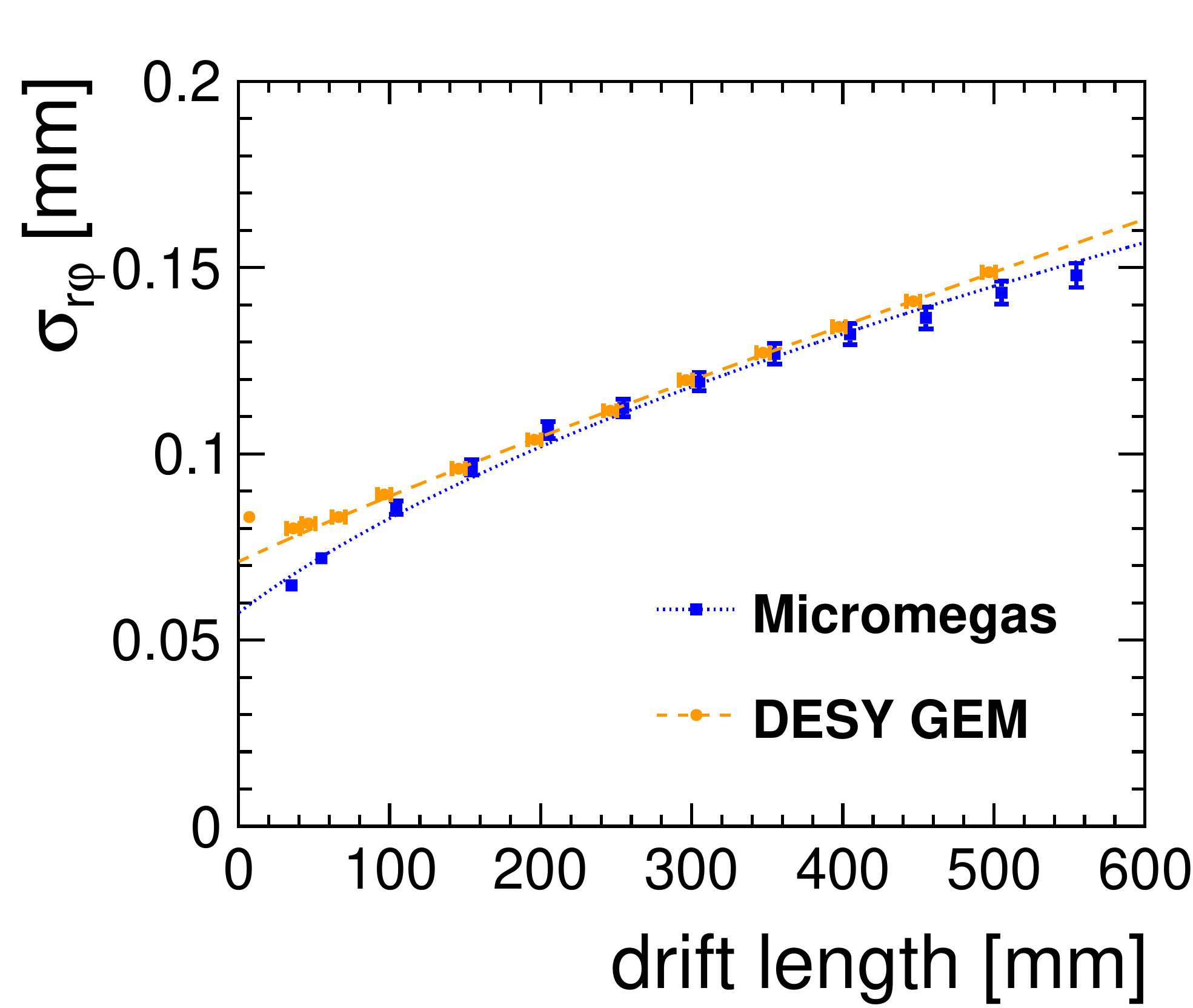}
\end{center}
\caption{Single-hit space-point $r-\phi$ transverse resolution 
plotted against drift distance for both GEM and Resistive Micromegas at 1.0 Tesla~\cite{ILD:paper}.}
\label{ilcmpgd:fig1}
\end{figure}

The LCTPC Collaboration strives to create an infrastructure for developing and testing new and
advanced detector technologies to be used at a future collider. The aim was to make possible
experimentation and analysis of data for institutes, which otherwise could not be realized due to lack
of resources. The LCTPC Collaboration welcomes participants from North-America.
The shared infrastructure comprises an analysis and software network; as well as instrumentation
setups for tracking detectors. A rather complete setup has been established at the DESY test beam, 
providing an environment for a world-wide effort in the development of a large TPC to be used as main tracking device at the ILC. 
It consists of the following items: 1) large scale (about 1 m) and
low mass field cage; 2) modular end plate system for large surface GEM and Micromegas systems;
3) MPGD detector modules; 4) prototype readout electronics; 5) magnet, 
supporting devices, HV, gas and cooling systems, and slow controls; 6) silicon envelope detectors; and
7) software developed within the MarlinTPC framework for simulation, calibration and
reconstruction of TPC data.
The LCTPC Large Prototype (LP) has a diameter of 720 mm and a length of 570 mm and fits into the 
1 Tesla superconducting
magnet PCMAG. Tracks measured within the LP can have up to 85 space points, using anode
pad readout, depending on the number of modules and pad size. The aim of these tests is not only to confirm the
anticipated single-point resolutions, but also to address issues related to the large size of this TPC,
like alignment, calibration, pulsed-electronics, cooling, electric and magnetic field distortions, dE/dx, and ion backflow.
Over the years analyses of data were performed from test beam measurements with the LP equipped with:
(i) DESY modules consisting of a triple GEM stack with standard CERN GEMs
that have a 50 $\mu$m thick kapton insulator and are wet-etched;
(ii) frameless Asian double GEM modules, which are stiffer since they are
made of 100 $\mu$m thick Liquid Crystal Polymer as insulator;
(iii) Micromegas modules with a micro-mesh stretched on a frame and
kept on top of a segmented anode with a thin layer of Diamond-Like Carbon deposited on
kapton used as a resistive material for charge dispersion.
Significant progress has been made in the manufacturing process of detection modules for all
readout options. Inhomogeneities of the electric field close to
the MPGD borders caused distortions of the recorded tracks. 
These distortions were corrected in upgraded modules and the residual distortions are treated
by common software package for both GEM and Micromegas pad-based readouts.
After alignment calibration and correction have been applied, the hit residuals line up around zero for both
technologies. In Figure \ref{ilcmpgd:fig1}, the measured transverse (xy-plane) space-point
resolutions are plotted as a function of the drift distance
for data collected in a 1 T magnetic field with GEM and Micromegas~\cite{ILD:paper}.
In all cases, the transverse resolutions were measured using the geometric mean of inclusive and exclusive residual 
distributions from track fits and fitted to the analytical 
form $\sigma_{\rm T}(z) = \sqrt{\sigma_{\rm T}^2(z=0) + D^2_T \, z}$, where $D_T$ is the coefficient of
transverse diffusion.  

\begin{figure}
\begin{center}
\includegraphics[angle=0,width=0.9\textwidth]{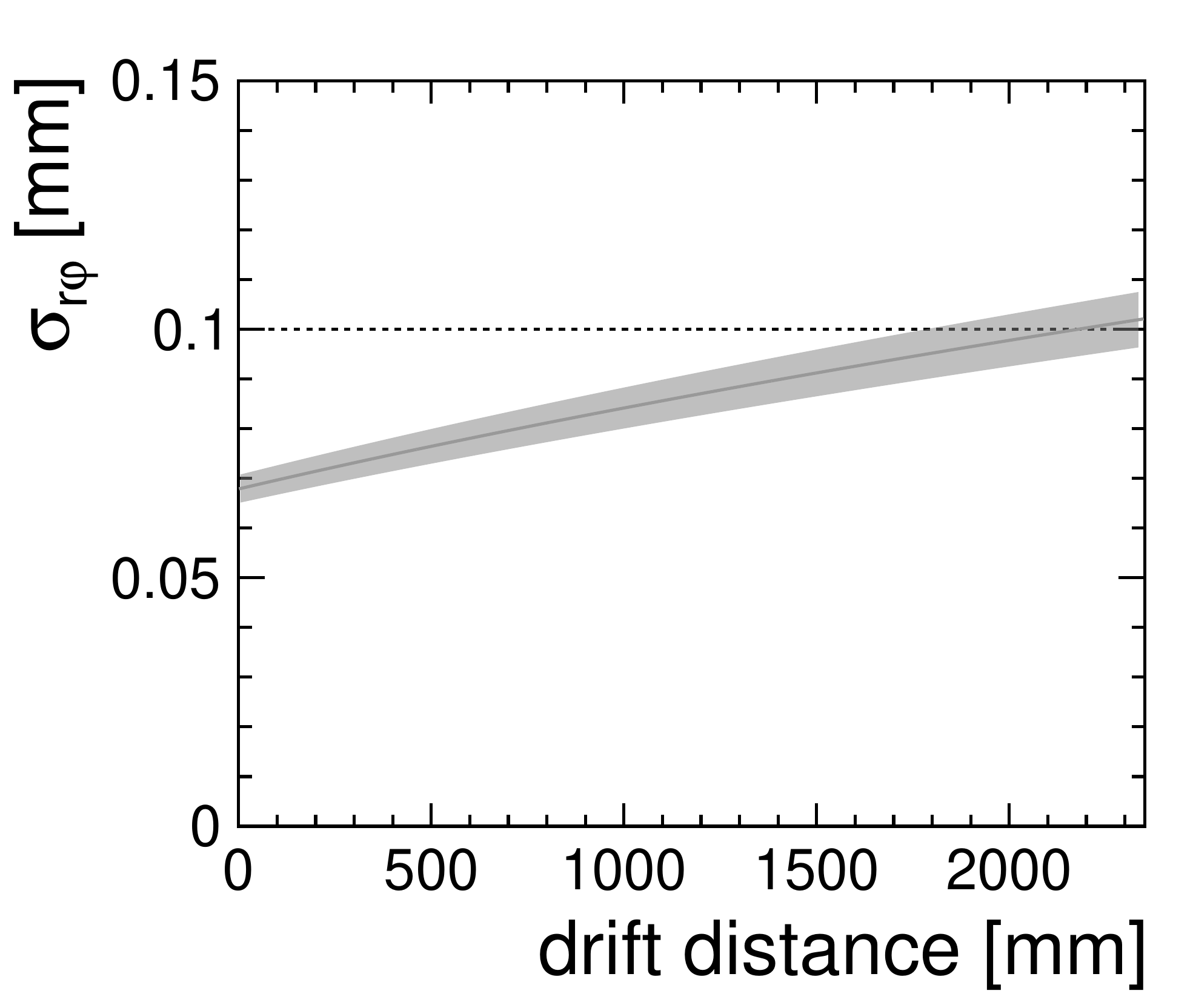}
\end{center}
\caption{Single-hit point $r-\phi$ transverse resolution extrapolation to a magnetic field of 3.5 T based on parameters measured with
the LCTPC prototype at 1.0 T and up to about 600 mm of drift distance. The resolution is plotted over the full ILD TPC length of 2.35 m including 1$\sigma$
band without any attachment for a perfectly controlled gas. The extrapolated resolution at an effective drift 
distance of 2.1 m in ILD is $<$ 100 $\mu$m, which
fulfils the designed TPC transverse resolution requirement. Adapted from Reference~\cite{ILC:detectors}.}
\label{ilcmpgd:fig2}
\end{figure}

Based on these results at 1 T with the LP, an extrapolation to the parameters of the ILD TPC has been done. 
The results are shown in Figure \ref{ilcmpgd:fig2}~\cite{ILC:detectors}. 
With a small attachment rate compatible with zero, the resolution requirement at the ILD experiment can be achieved. 
TPCs in running experiments as T2K~\cite{T2K:TPC} or ALICE~\cite{ALICE:TPC} demonstrated that the necessary control of the gas conditions is possible.
The expected single-point hit resolution in a magnetic field of 3.5 T confirms that pad-based GEM and Micromegas technologies
meet the requirements of the proposed ILD-TPC for the future ILC, which is single-hit resolutions 
of $\sigma_{r\phi}(z=0) \sim 60$ $\mu$m and
$\sigma_{r\phi} < 100 \mu$m in the transverse plane for all tracks after 2.1 m of drift. 
Similar measurement and extrapolation on longitudinal $z$, or time, single-hit resolution show that the ILD TPC requirements can be also achieved~\cite{ILC:detectors}.

\begin{figure}
\begin{center}
\includegraphics[angle=0,width=0.8\textwidth]{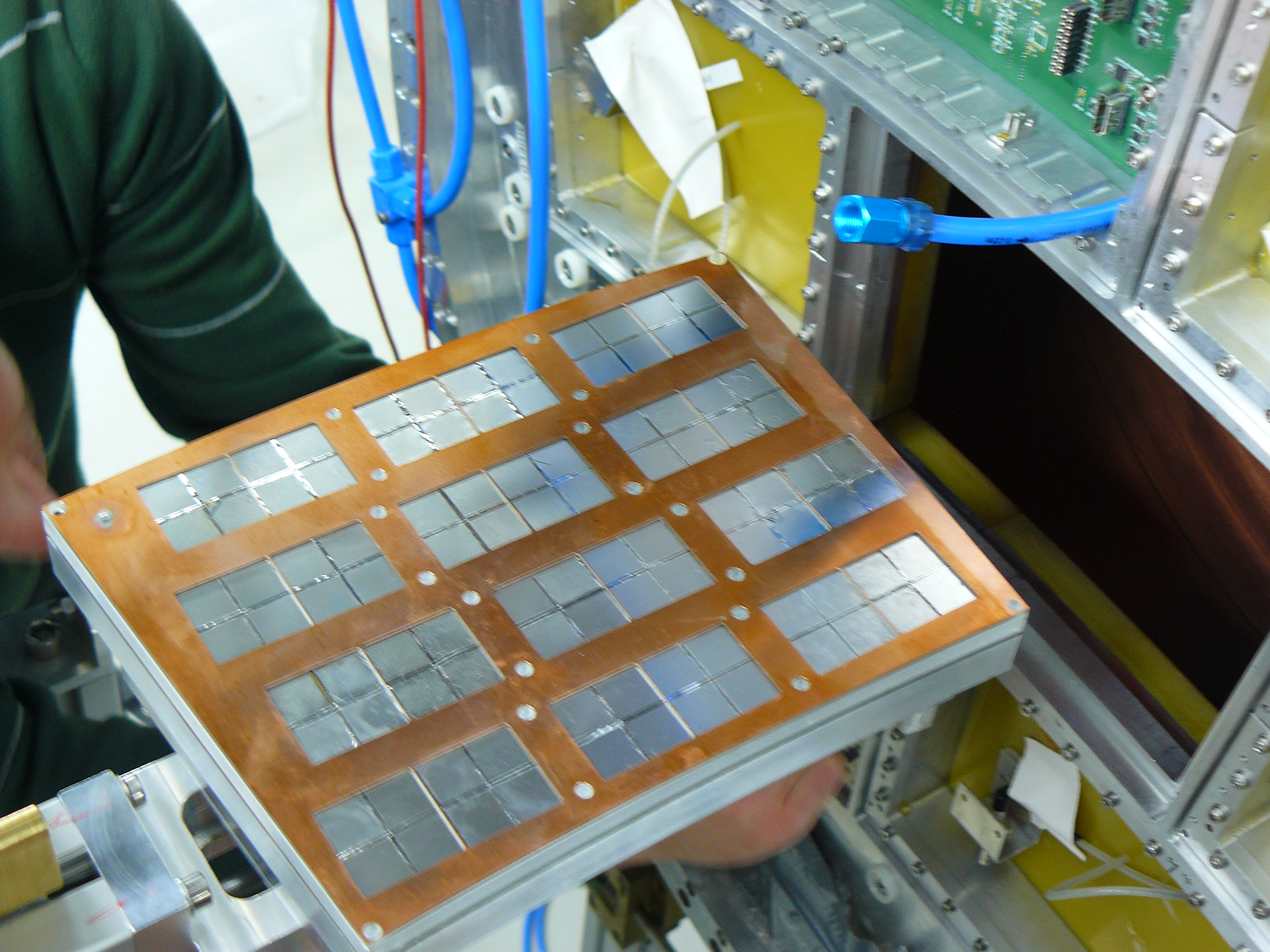}
\includegraphics[angle=0,width=1.0\textwidth]{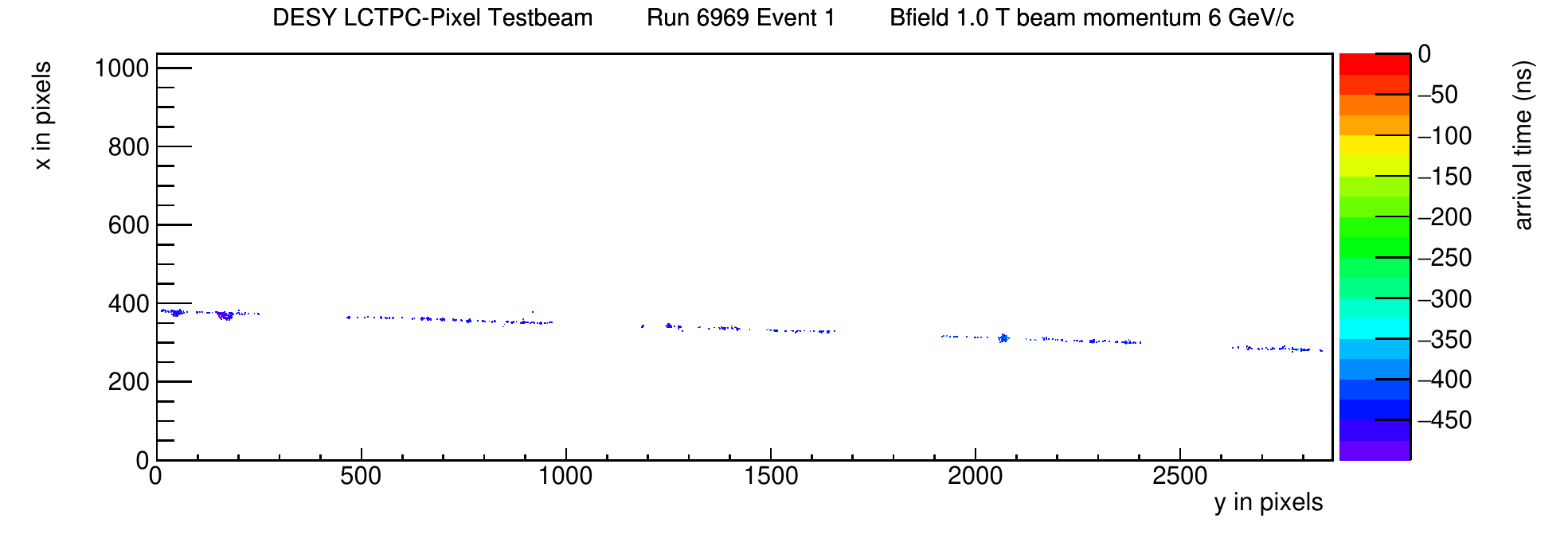}
\end{center}
\caption{Fully equipped LP module with 96 GridPix detectors being mounted in the LP and a GridPix event display.}
\label{ilcmpgd:fig3}
\end{figure}

Figure \ref{ilcmpgd:fig3} shows a fully equipped LP module with 96 GridPix detectors ($\sim$2 cm$^2$ cell) being mounted in the LP.
The readout structure consists of 160 GridPixes with a total of 10.5 million pixels, each of a size of 55 $\mu$m $\times$ 55 $\mu$m.
A GridPix event is depicted in Figure \ref{ilcmpgd:fig3}.
Preliminary results on the performance of GridPix readout shows great promise. Single electron diffusion measurements for GridPix
points to similar, or better, performance as obtained with GEM or Resistive Micromegas for both space-point resolution and dE/dx PID. 
This indicates that the GridPix novel technology has great potential and is worth further investment in R\&D~\cite{TIPP2021}.

One critical issue concerns potential field distortions due to ion accumulation within the drift volume of the chamber. 
At ILC, this can be mitigated by implementing an ion 
gating between bunch trains, using large aperture GEM foils. During bunch trains, the
voltage difference between the GEM sides is configured to allow drift electrons cross the GEM and
reach the amplification region. Outside bunch trains, the voltage difference is reversed so that ions
produced in the gas amplification region stay confined and are guided to the GEM surface where they
are absorbed. This GEM ion gating system has been assembled and tested. It is designed to operate on top 
of both a triple-GEM or a resistive Micromegas. The electron transparency of the GEM gating has
been determined with different measurements and corresponds to 82\% as expected from
simulations. The ion blocking power is deemed adequate, but still has to be further elaborated and quantified~\cite{ILD:paper,ILC:detectors}.
First measurements have been initiated with a fast HV switching circuit that has to be established
and tested in B-field of 3.5-4 T. New electronics for R\&D purposes based on the programmable ASICs 
is being developed within the LCTPC Collaboration.  Despite the pulsed mode of data taking with power-pulsing, the readout electronics
and the endplates will require a cooling system.  A fully integrated solution has been already tested on seven modules during a testbeam. This 
two-phase CO2 cooling is a very promising candidate.
The new work on modules, update ion gating, cooling and electronics are to consolidate and
improve the already proven result that a MPGD TPC can meet the ILD requirement for physics exploitation at the ILC.

The three options for the ILD TPC  under consideration for the MPGD signal amplification and readout are~\cite{TIPP2021}:
\begin{itemize}
\item GEM: the ionization signal is amplified by passing through a multi-layered structure with avalanche in the holes of the GEM foil and the charges collected on pads.
\item Resistive Micromegas: the ionization signal is amplified between a mesh and the pad plane. The charge is induced on the pads under a resistive coating.
\item GridPix: the ionization signal is amplified as for the Micromegas case, but collected on a fine array of silicon pixels providing individual pixel charge and timing using Timepix3 
ASICs (Timepix4 is being studied for future application).
\end{itemize}

For the GEM and resistive Micromegas options, the typical pad sizes are a few mm$^2$ and spatial resolution is improved by combining the track signals of several adjacent pads. For the GridPix option,
the pixel size of 55 microns matches the pitch of the mesh holes, providing pixel sensitivity to single ionization electrons. The GridPix spatial resolution is improved and the dE/dx PID enhanced by counting pulse-heights and clusters. Overall, the LCTPC is conceptually ready as it meets design specifications and is engineeringly possible.
Here, the single-hit transverse resolutions were shown from multiple testbeam campaigns at 1 T magnetic field and the extrapolated result 
at 3.5 T clearly demonstrate that single point resolution of 100 $\mu$m over about 220 points is achievable with MPGD technology (GEM, MM or GridPix).
This translates from simulation to the desired two-hit separation of $\sim$2mm and a momentum resolution $\delta(1/p_T) \simeq 10^{-4} /$ GeV/c (at 3.5 T), which is 
the required performance of the TPC as a standalone tracker at ILD for ILC. Measurements on dE/dx and ion gating also confirm the needed performance at ILC.
Those results and conclusions are based on an established experimental R\&D program.

The framework set by the LCTPC international Collaboration and with the International Development Team (IDT) for the realization of the ILC in Japan offer a window
of opportunities for a significant engagement of the LCTPC American observer institutions with dedicated funding for advancement in instrumentation. MPGD TPC technologies can 
be linked with
other application domains where MPGD are bringing benefits and synergy with other areas of applied physics reported at the Instrumentation Frontier that relies on 
microscopic structure devices.  North America has infrastructures in place at national laboratories and universities to enhance and guide the development of 
instrumentation based on micro pattern detectors for ILC.


\section{Belle II Upgrade}

Intensity frontier experiments, particularly those at the $B$ Factories, require high-precision tracking for fairly low-momentum tracks in the presence of high event and beam-induced background rates~\cite{Belle:detector}. Such experiments typically rely on drift chambers to measure helical segments of charged tracks with a minimal material budget. Operational experience in the early stages of Belle II, the current state-of-the-art $B$ Factory experiment, has shown that the drift chamber technology may be reaching its limit due to high occupancy. To address this issue, the Belle II  Collaboration has developed and simulated a first conceptual design for a TPC-based tracking system for a hypothetical future 
ultra-high luminosity $B$ factory experiment. For convenience and concreteness, it is supposed that the SuperKEKB and Belle II upgraded accelerator and detector will operate with the same beam energies but with five times the maximum instantaneous luminosity ($5.0 - 6.5\times 10^{35} {\rm{cm}^{-2} \rm{s}^{-1}}$), and it is assumed that the proposed tracking system is surrounded by existing Belle II components. However, in principle the concept is equally suitable for a Belle II upgrade or a future unrelated intensity-frontier experiment. 

In the ultra-high luminosity Belle II upgrade scenario, the geometry is constrained by the existing PID and electromagnetic calorimetry systems. With this constraint, three competing proposals were considered. The first is an upgraded drift chamber. This solution is challenged and almost unsuitable given the current challenges of operating Belle II's drift chamber (CDC) at SuperKEKB current instantaneous luminosity. The second option is a full silicon tracker. Preliminary simulations of such a system is found to significantly degrade the $p_T$ resolution of tracks due to increased multiple scattering. This, coupled with the intrinsic cost and structural difficulties of such a system, suggests that such a system might not be suitable. The third option is a TPC-based tracker, which should provide a significant reduction in occupancy because it is a true three-dimensional detector, while drift chambers are effectively two-dimensional. However, a TPC tracker has some intrinsic limitations: it cannot provide a trigger signal as the CDC does, and event pileup can be very high due to the long electron drift time. Other questions raised by the tracking TPC concept, include (i) reliable association of tracks with unique events despite a high degree of event overlap (ii) beam-induced background hits (iii) ion backflow mitigation with possible continuous readout design (without gating) at high physics event rates. Overall, can a TPC match the tracking performance of a CDC by using a high number of space-time points to overcome the limitations of diffusion? These problems can be overcome based on a conceptual design that addresses such challenges. 

For the preliminary results presented here, the current Belle II CDC is replaced by a TPC with a single drift volume and readout on the backward endcap. Second, the current silicon vertex detector (VXD) is replaced with a new detector, which is based on a so-called VTX upgrade proposal. In order to maintain an annular cylinder geometry for the TPC, the VTX is extended to a radius of 44 cm. Third, a multilayer fast timing detector, possibly silicon, placed at $r=25$ cm or $r=45$ cm is used in order to replace the triggering role of the CDC and additionally provide particle identification (PID) via time-of-flight (TOF) for low-$p_{T}$ tracks.

In order to focus on these key challenges instead of more basic technical design optimizations, Belle II preliminary studies borrow heavily from work already done by the LCTPC Collaboration for 
the ILD TPC~\cite{Belle:GridPix}. The basic conceptual design consists of a single gas volume of length 242 cm with high-resolution readout tiling the backward endcap without ion gating with continuous data taking mode. 
It is assumed that the TPC uses atmospheric pressure T2K gas with a drift field ($289$~V/cm) that minimizes the drift time of electrons ($v_D= 7.89$ cm$/ \mu$s, leading to a maximum drift time of roughly 30 $\mu$s for a maximum drift length of 242 cm). From the simulation, the longitudinal and transverse diffusion coefficients 
are $\sigma_L=$ 200 $\mu$m$/ \sqrt{\rm{cm}}$ and $\sigma_T=$ 84 $\mu$m$/ \sqrt{\rm{cm}}$ within a B-field of 1.5 T. For charge amplification and readout, the GridPix system proposed for use in LCTPC is used with an array of $55 \mu\rm{m} \times 55 \mu\rm{m}$ pixels with a Micromegas mesh mounted onto the surface. As for the ILC TPC, this technology presents a number of advantages: first, the small pixels and direct mapping between amplification cells and pixels constitute essentially a best-case resolution scenario. Second, in theory such a sensor can be operated in binary readout mode in which each individual hit represents exactly one electron and consists only of the pixel ID and a threshold-crossing time ID. This can dramatically reduce the data throughput, which is anticipated to be a significant technical challenge at ultrahigh luminosities and with continuous readout. Thirdly, the total number of ions produced during amplification can be far smaller, perhaps leading to a reduction in the number of backflowing ions. Finally, it is relatively easy to implement such a detector in the actual Belle II digitization simulation. 

\begin{figure}
	\centering
	\includegraphics[width=0.9\textwidth]{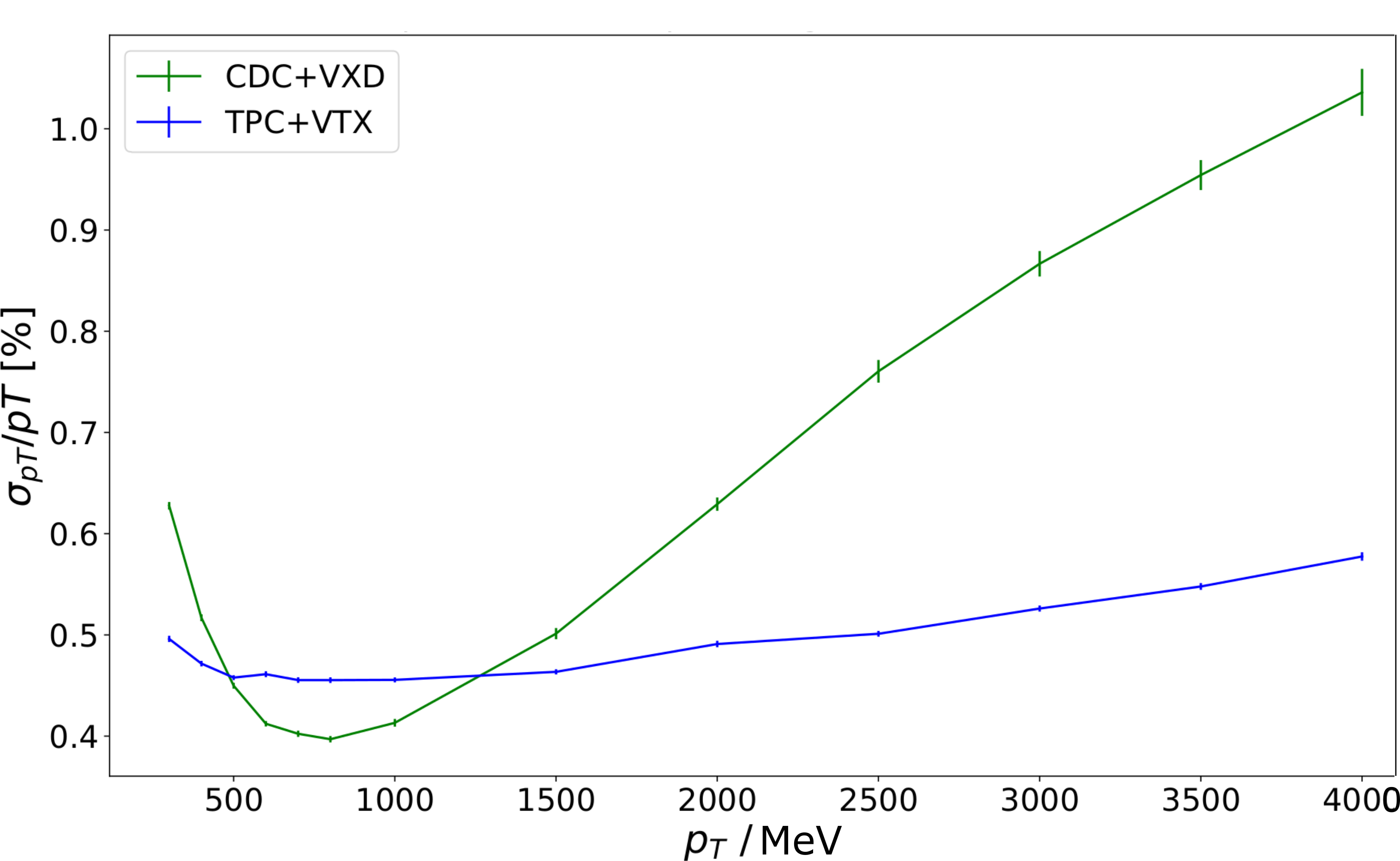}
	\caption{A comparison of the average $p_T$ resolution for the CDC+VXD and TPC+VTX tracking systems.}
	\label{fig:pTRes_PTCol}
\end{figure}

The key question concerning tracking performance is whether the increase in the number of spatial hit points (up to the theoretical maximum of number of ionizations) can win out over thermal diffusion 
when determining tracking parameters. Here the focus of the result presented is primarily on the transverse momentum $p_T$. A set of simulated muons in discrete bins of $p_T$ and distributed 
uniformly in $\theta$ over the acceptance of the TPC.  Figure \ref{fig:pTRes_PTCol} shows the simulated $p_T$ resolution for the current (CDC+VXD) versus the proposed (TPC+VTX) tracking systems averaged over all track polar angles $\theta$. The vertical offset is due to multiple scattering, which is largely determined by the material budget of the vertex detectors, which for the VTX is highly speculative. In principle, due to significant thinning of each layer, the TPC+VTX should be able to achieve comparable or lower levels of multiple scattering compared to the CDC+VXD.
The linear slope in $p_T$, due to position measurement resolution, is far shallower for the TPC compared to the CDC. This is due to the large number of hit points in the TPC, rendering its effective point resolution far superior to the hit resolution in the CDC. In conclusion, the simulated TPC result shows that it is possible to match and even surpass the tracking resolution of the CDC. However, the tracking performance in the critical range $p_T< 1$ GeV depends more strongly on the amount of material in the VTX than it does on the differences between the CDC and TPC.  
It is also found that binary readout with relatively larger pixels ($200 \mu\rm{m} \times 200 \mu\rm{m}$) is also sufficient to meet the performance objectives of the tracking system 
of ultra-high luminosity Belle II upgrade, which decreases the channel count and data throughput of the system, perhaps offsetting some of the costs. 

Based on these studies, a gas TPC-based tracking system seems viable for an intensity frontier experiment like the hypothetical ultra-high luminosity Belle II upgrade. This conceptual design relies heavily on the capabilities of the 
GridPix sensor, particularly the association of a single pixel with a single electron with low ion backflow and excellent 3D resolution. The primary difficulty of such a system is the tiling of these sensors 
over the $> 3 \rm{m}^2$ endplate. However, it offers an amazing opportunity for groups in the US interested in R\&D for MPGD for future upgrade with the Belle II detector.


\section{TPC for CEPC}

The Circular Electron Positron Collider (CEPC) has been proposed as a Higgs/Z factory in 
China~\cite{CEPCStudyGroup:2018ghi}. The baseline design of a CEPC detector consists of a tracking system composed of a vertex detector with three concentric double-sided pixel layers, a high precision (about 100 $\mu$m) large volume TPC and a silicon tracker in both barrel and end-cap regions. The tracking system has similar performance requirements as for the ILD detector, but without power-pulsing, which leads to additional constraints on detector specifications, especially for the case of the machine operating at Z-pole energy with high luminosity. Until a decision on a tracker for a future circular collider in China can be reached, 
a number of tasks are still remaining regarding the TPC research. Such tasks include the full simulations of the TPC performance in the CEPC environment, further design of the low power consumption readout electronics, UV laser calibration methods and cooling options~\cite{Yuan:2021sgq}.
Some of the key challenges to be addressed in the near future are the physics requirements for the TPC performance towards the
inclusive CEPC physics program. MPGD technology, though quite far advanced in some aspects, still needs a significant effort from key partners.
Nonetheless, the CEPC TPC requirements and challenges for the detector are similar than the ones described for the ILD, and thus achievable with existing MPGD technologies. R\&D activities are actively ongoing in China and could potentially lead to partnership with the USA.

Overall, the TPC at CEPC has been inspired by the ILC-TPC development.
Contrary to the ILC, Z-pole running at CEPC with luminosity of about 10$^{36}$ cm$^{-2}$ s$^{-1}$ prevents gating mode TPC operation. 
Hence, a hybrid gaseous structure~\cite{Yuan:2021sgq} and a double micro-mesh gaseous structure~\cite{ZHANG2020161978} were proposed and studied in depth, and the key factor of the gain times the IBF suppression ratio shows good promise.
In this situation, GridPix is also an attractive option, which provides the
high granularity needed to resolve individual electron clusters and to determine energy
loss by the cluster counting technique. 
The CEPC TPC requires transverse ($r-\phi$) single-hit space-point resolutions of less than 100~$\mu$m and longitudinal ($z$) time resolution of about 100 ns.   The physics 
goals require dE/dx resolution of less than 5\% with an even better particle identification separation with cluster counting.
Most conditions set by the CEPC tracking systems can be met by MPGDs as proven by LCTPC effort. 
Such detector development offers a possibility for partnership between the LCTPC collaboration groups and China.

\section{Other Applications and Synergy}

In the previous sections, the focus was on MPGDs for (i) the International Large Detector (ILD) at the International Linear Collider (ILC), (ii) the Belle II upgrade at SuperKEK B-Factory, and (iii) the TPC for a detector at the Circular Electron Positron Collider (CEPC). 
Despite the mention of only those initiatives, radiation detection through the ionization in micro-pattern gas amplification devices has many fields of applications ranging from particle, nuclear 
and astro-particle physics experiments with and without accelerators. MPGDs are applicable as well in medical imaging and 
homeland security screening.  Several new micro-pattern gas amplification concepts, such Thick-GEMs (THGEM) or patterned resistive-plate devices, are also under study for 
calorimeter and muon systems. 
To name only two examples of synergy between TPC-tracking and other instrumentation development, particle-flow algorithm from continuous detection in MPGD devices promises to deliver 
calorimetry excellent jet energy resolution, while large sensitive area MPGDs serve as natural logical choices for muon systems at future multi-purpose 4$\pi$ spectrometers.

In some tracking applications coarse patterned readout can be used for experiments requiring very large-area coverage with moderate spatial resolutions. 
Such conceptual design of the newest micro-pattern devices are in fact quite suitable 
for production and development with industrial partners as it was proven for recent deployment of LHC detector upgrades from Run3 onwards. 
MPGDs have indeed been proven to be a natural choice of technologies for large sensitive area for muon systems as demonstrated by CMS and ATLAS who recently upgraded part of their muon systems. 
CMS opted for GEM; while ATLAS used Micromegas staggered with small-strip Thin Gap Chamber (sTGC). Both CMS and ATLAS have numerous institutions from the United States funded by DOE
or NSF who participate in the deployment of the upgraded MPGD-based muon detectors.  
There have been major recent MPGD developments
for ATLAS~\cite{ATLAS:NSW} and CMS~\cite{CMS:GEM} muon system upgrades (from Run 3 onwards) that established design concepts and technology goals; while addressing engineering and integration challenges.
The ATLAS resistive Micromegas are set to suppress destructive
sparks in the high rate environments, while the CMS GEM single-mask with self-stretching techniques
enable the reliable production of large-size foils and significantly reduce detector assembly time.
For example, (i) the completion of the ATLAS New Small Wheels for Run3 relied on the expertise
of the TRILAB company in Nevada for the precise machining, etching and pressing of anodes boards of meter-scale for sTGC that share and employ very similar attributes 
as MPGD readouts; while (ii) groups from the US participated in the GEM deployment of the stations GE1/1 and GE2/1 to complement Cathode Strip Chamber (CSC) in the CMS Muon endcaps.
MPGDs are foreseen as a technology of choice for the future upgrades at HL-LHC operation from 2025 onwards.
Indeed, MPGDs are planned for further upgrades of the muon
systems of CMS and ATLAS based on GEMs with high granularity spatial segmentation and small-pad resistive Micromegas, respectively.
The development of fabrication techniques 
of MPGDs for LHC upgrades towards HL-LHC showed that large-scale applications are possible from design to deployment with a cost-effective manufacturing where the US can potentially 
play a role. 

Several groups worldwide, who often collaborate with US groups, have developed the ability to either produce large PCB boards or stretch large-area meshes for the 
construction of MPGD devices.  
Researchers at the Florida Institute of Technology in Melbourne, Florida, USA, are developing, under 
a grant from the U.S. Department of Homeland Security, GEM detectors that utilize cosmic ray muons for homeland security. The readout electronics needed for MPGDs share many common attributes 
with the electronics for silicon detectors; so this this should be kept in mind when taking a global look at detector development. 
MPGDs can also fulfill the stringent experimental constraints imposed by future nuclear, hadron physics experiments, and heavy ion facilities. Another example would be GEM, Micromegas or GridPix
operating at the Electron-Ion Collider (EIC), offering intrinsic high-rate capability ($10^6$ Hz $/\rm{mm}^2$), spatial resolution (down to 30 $\mu$m), multi-particle
resolution ($\sim$ 500 $\mu$m), and superior radiation hardness. Although normally
used as planar detectors, MPGDs can be bent to form cylindrically curved ultra-light inner tracking systems with an amazing PID capability, without support and cooling structures. 
Again, those attributes show how effective MPGDs can be on flagship projects driven by the US P5 community.







\end{document}